\documentclass[conference]{IEEEtran}
\IEEEoverridecommandlockouts

\usepackage{cite}
\usepackage{amsmath,amssymb,amsfonts}
\usepackage{algorithmic}
\usepackage{graphicx}
\usepackage{textcomp}
\usepackage{xcolor}
\usepackage{xspace}
\usepackage{booktabs}
\usepackage[hyphens]{url}
\usepackage{hyperref}
\usepackage{comment}
\usepackage{multirow}
\usepackage{multicol}
\usepackage{balance}

\begin{document}

\newcommand{\tool}{{\textsc{AutoPRTitle}}\xspace}
\newcommand{\pr}{{pull request}\xspace}

\makeatletter
\newcommand{\newlineauthors}{%
  \end{@IEEEauthorhalign}\hfill\mbox{}\par
  \mbox{}\hfill\begin{@IEEEauthorhalign}
}
\makeatother

\title{\tool: A Tool for Automatic Pull Request Title Generation}
\author{\IEEEauthorblockN{Ivana Clairine Irsan$^*$, Ting Zhang$^*$, Ferdian Thung, David Lo and Lingxiao Jiang}
\IEEEauthorblockA{School of Computing and Information Systems, Singapore Management University\\
Email:
\{ivanairsan,\:tingzhang.2019,\:ferdianthung,\:davidlo,\:lxjiang\}@smu.edu.sg
}}




\maketitle

\begingroup\renewcommand\thefootnote{*}
\footnotetext{Both authors contributed equally to this research.}
\endgroup

\begin{abstract}
With the rise of the \pr mechanism in software development, the quality of pull requests has gained more attention.
Prior works focus on improving the quality of \pr descriptions and several approaches have been proposed to automatically generate \pr descriptions.
As an essential component of a \pr, \pr titles have not received a similar level of attention.
To further facilitate automation
in software development and to help developers draft high-quality pull request titles, we introduce \tool.
\tool is specifically designed to generate \pr titles automatically.
\tool can generate a precise and succinct \pr title based on the \pr description, commit messages, and the associated issue titles.
\tool is built upon a state-of-the-art text summarization model, BART, which has been pre-trained on large-scale English corpora.
We further fine-tuned BART in a \pr dataset containing high-quality \pr titles.
We implemented \tool as a stand-alone web application.
We conducted two sets of evaluations: one concerning the model accuracy and the other concerning the tool usability.
For model accuracy, BART outperforms the best baseline by 24.6\%, 40.5\%, and 23.3\%, respectively.
For tool usability, the evaluators consider our tool as easy-to-use and useful when creating a \pr title of good quality.

Source code: \url{https://github.com/soarsmu/Auto-PR-Title}.

Video demo: \url{https://tinyurl.com/AutoPRTitle}.

\end{abstract}


\section{Introduction}
Collaborative coding platforms, such as GitHub, widely adopt the \pr mechanism.
Developers make a local copy from the main repository and make changes in their local copy (i.e., branches).
After they are satisfied with their changes, they can request to merge their branch to the main repository by opening a new \pr.
A \pr consists of a \pr title, a \pr description (optional), and several commits.
Similar to how an issue quality is important for bug triaging, the quality of a \pr also makes an impact on the decision of whether a \pr gets merged~\cite{yu2015wait}.
As an essential component of a \pr, the title has received less research attention compared to the description.
Several prior works~\cite{liu2019automatic,fang2022prhan} have proposed different approaches to generate \pr descriptions based on the information from the commits, i.e., commit messages and comments in code changes.

In our earlier work~\cite{prtiger2022-arxiv}, as the first work on \pr title generation, we built a dataset named PRTiger containing pull requests from 495 popular GitHub repositories.
We formulate the \pr title generation task as a one-sentence summarization task.
The source sequence is the concatenation of the \pr description (if any), commit messages, and the related issue titles (if any).
The target sequence is the \pr title.
We utilized BART~\cite{lewis2019bart}, which is a pre-trained sequence-to-sequence model that has achieved a remarkable performance in several summarization tasks. It achieved the best performance among the approaches that we evaluated.

In this paper, we present \tool, which is a tool for automatically generating \pr titles. This is a demonstration paper accompanying our above-mentioned full research paper~\cite{prtiger2022-arxiv}. 
We implemented \tool as a web application.
\tool takes a new \pr link, a \pr description, and related issue links from the user as inputs.
\tool can extract the commit messages from the given \pr link.
If a related issue link is provided, \tool will also extract the issue title.
The commit messages, the issue title, and the \pr description are then fed to the BART model. 
\tool will then provide a \pr title suggestion.

To evaluate our tool, including the underlying model, we performed an automatic evaluation to measure model accuracy, and a manual evaluation to measure the tool usability.
We compared BART with existing approaches for software artifact generation.
We consider two approaches for similar tasks as the baselines, i.e., PRSummarizer for \pr description generation~\cite{liu2019automatic}, and iTAPE for issue title generation~\cite{chen2020stay}.
The experimental results from the automatic evaluation show that the fine-tuned BART~\cite{lewis2019bart} produced the best performance. It achieves the highest ROUGE-1, ROUGE-2, and ROUGE-L F1-scores of 47.22, 25.27, and 43.12, which outperform the best baseline by 24.6\%, 40.5\%, and 23.3\%, respectively.
For the manual evaluation, we asked 7 evaluators to rate our tool. They consider our tool to be easy-to-use and useful when drafting a good \pr title.
\section{Compared Methods}
As pre-trained models have produced remarkable performances in many tasks (e.g.,\cite{zt2022benchmark,zhang2020sentiment}), we utilize BART (\texttt{facebook/bart-base}), which has demonstrated its ability to solve text summarization tasks in the natural language processing field~\cite{lewis2019bart}.
We also presented some existing approaches that solve similar tasks, i.e., PRSummarizer~\cite{liu2019automatic} for \pr description generation and iTAPE~\cite{chen2020stay} for issue title generation as our baseline methods.
We briefly describe these approaches as follows:

\begin{itemize}
    \item{\textit{BART}~\cite{lewis2019bart} adopts a standard Transformer~\cite{vaswani2017attention} architecture. BART was pre-trained by (1) corrupting the source sequence with noise functions and then (2) learning to reconstruct the original text. It was pre-trained in the same corpora as RoBERTa~\cite{liu2019roberta}. The pre-training corpora cover 160GB of text, including news, books, stories, and web text. BART achieved state-of-the-art performance in several text generation tasks, such as dialogue response generation on \textsc{CONVAI2}~\cite{dinan2020second}}.
    \item{\textit{PRSummarizer}~\cite{liu2019automatic} leverages the pointer generator~\cite{see2017get} to handle the out-of-vocabulary (OOV) issue. Besides, to mitigate the gap between the loss function and the evaluation metrics, i.e, ROUGE scores, PRSummarizer adopts reinforcement learning to optimize the performance in terms of ROUGE scores directly.}
    \item{\textit{iTAPE}~\cite{chen2020stay} was originally proposed to generate issue titles. The source sequence is the issue description, and the target sequence is the issue title. iTAPE leverages a sequence-to-sequence model. To cope with the OOV issue, it combines two techniques, i.e., a lightweight tagging method and a copy mechanism~\cite{gu2016incorporating} with a pointer generator}.
\end{itemize}
\section{Tool Architecture}
\subsection{Overview}
\label{subsection:overview}

\begin{figure}[t]
\includegraphics[width=\linewidth]{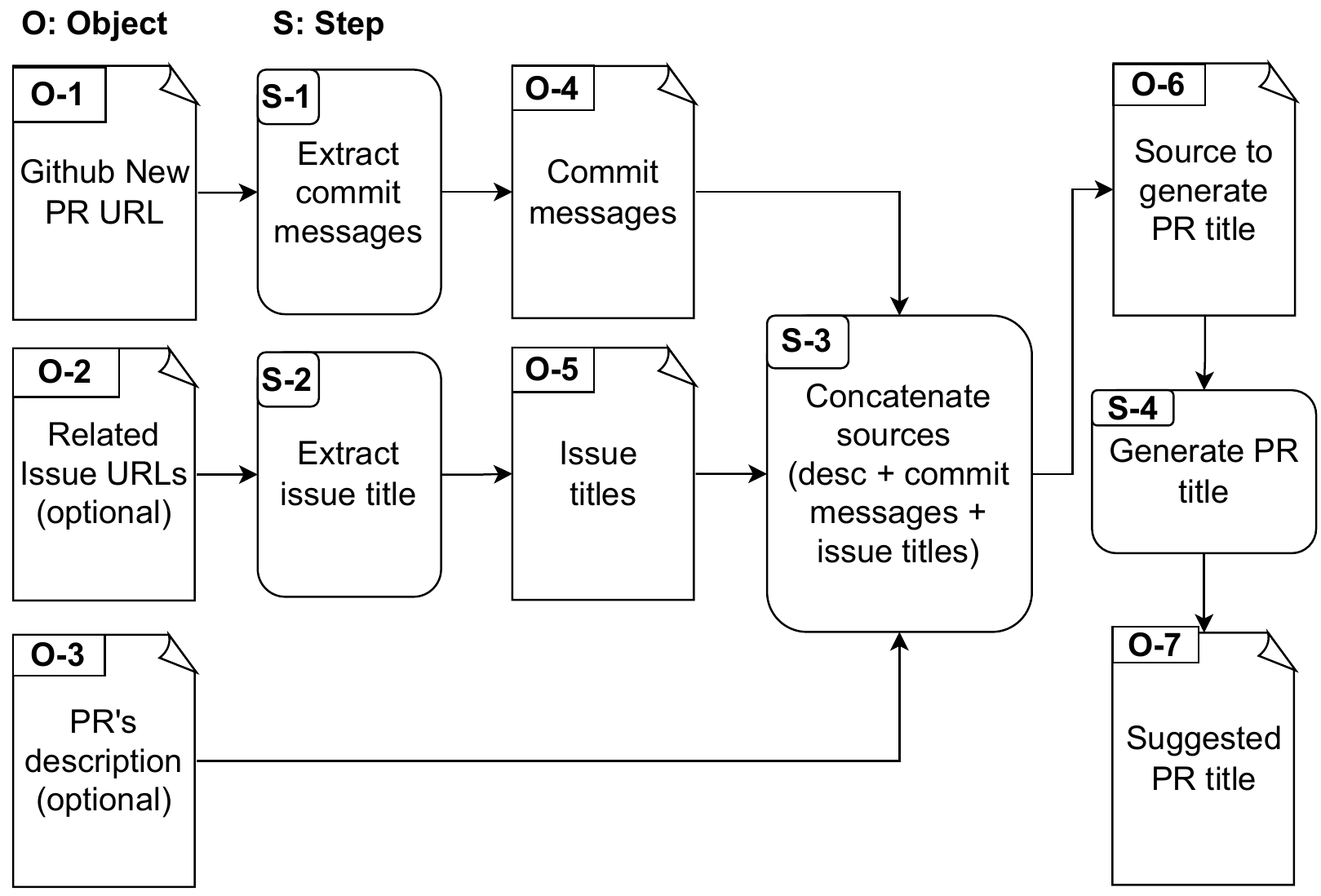}
\caption{The architecture of \tool}
\label{fig:flow-architecture}
\end{figure}

The architecture of \tool is presented in Figure~\ref{fig:flow-architecture}.
\tool is built upon simplicity.
It is packaged as a web application that could be easily deployed using a Docker container.
This tool has three input sources to generate a \pr title suggestion: a new pull request URL (O-1), related issues' URLs (O-2), and a \pr description (O-3).
The new pull request URL (O-1) is a mandatory input, as it is needed to extract (S-1) the commit messages (O-4).
On the other hand, related issue URLs (O-2) are not mandatory as a pull request is not necessarily a fix to an issue.
If related issue URLs are given, \tool will extract the corresponding issue titles (S-2, O-5).
While a \pr description (O-3) is an optional input, users are encouraged to fill it in to get a better-generated title.
All of the three sources are then concatenated (S-3) before being used as a textual source sequence (O-6) to generate a suggested \pr title (S-4, O-7).

\begin{figure}[t]
\includegraphics[width=\linewidth]{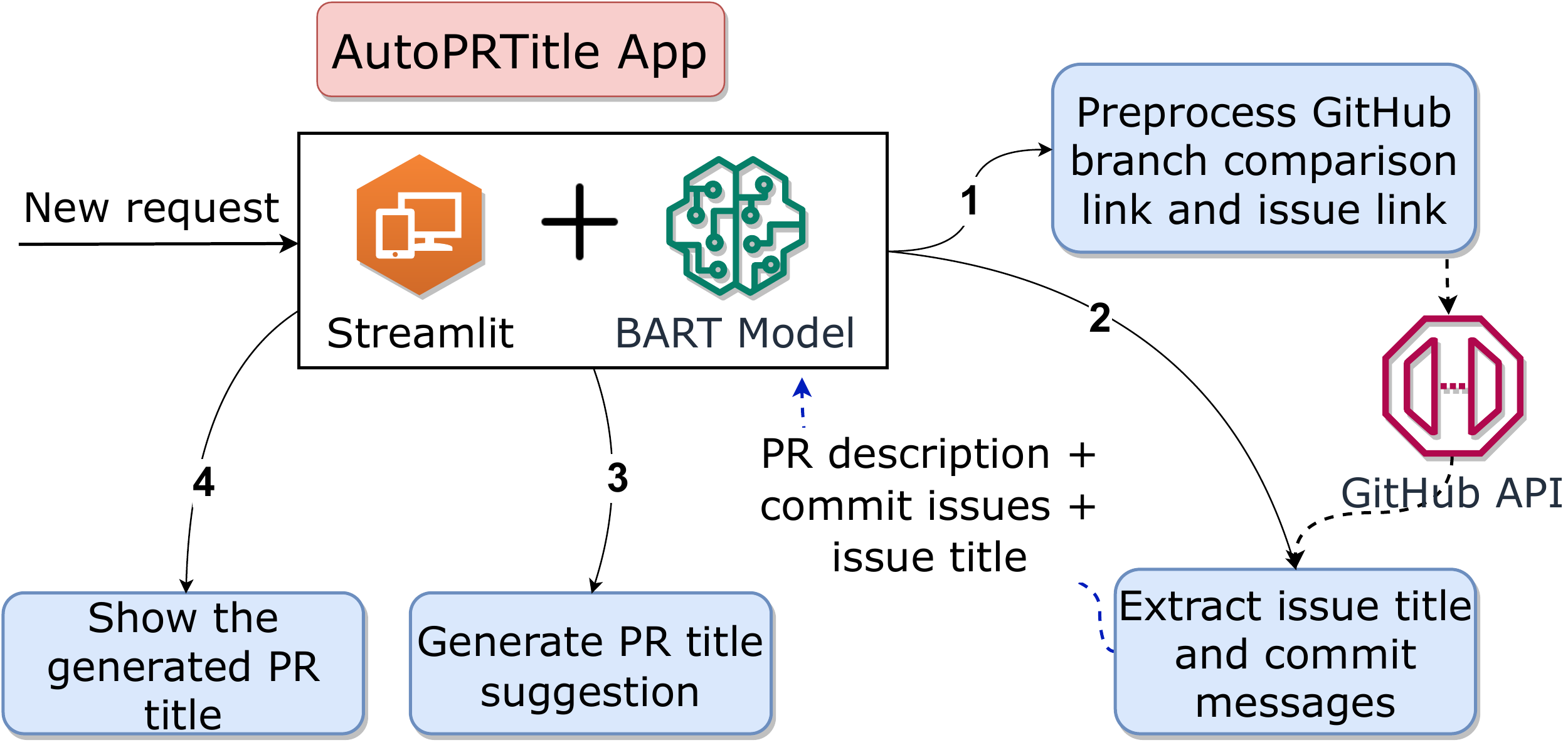}
\centering
\caption{The workflow of \tool}
\label{fig:flow-app}
\end{figure}

\subsection{Implementation Details}
Figure~\ref{fig:flow-app} shows the detailed workflow of \tool. \tool is developed on top of \texttt{Streamlit}~\cite{Streamlit}, an open-source app framework that is built as a Python3 library.
It enables us to integrate the fine-tuned BART model with a web interface seamlessly.
\tool uses a BART Model as the underlying model to generate the pull request title. The BART Model is loaded once at start up, and will be called every time there is a request to generate a \pr title.

\textbf{First}, \tool preprocesses the given new \pr link (e.g., \url{https://github.com/microsoft/vscode/compare/main...TylerLeonhardt/copy-after-action}) and issue links.
For both links, \tool then replaces the ``github.com" in the link with ``api.github.com/repos" to create a GitHub API's request URL.
After the GitHub API's request URLs are obtained, \tool makes REST API calls to these URLs.

\textbf{Second}, \tool extracts the issue title and commit messages from the GitHub API response.
The response associated with the GitHub branch comparison link contains a list of commits ready to be merged to the target branch. 
We extract commit messages from these commits' details. For the response associated with the issue link, the corresponding issue details are returned. 
We will extract the issue title from these issue details. At this point, as no pre-processing step is done toward the \pr description, all the input sources are ready to be concatenated. \tool proceeds to concatenate the \pr description, the commit messages, and the issue titles.

\textbf{Third}, using the BART Model, \tool generates the \pr title based on the concatenated text. While the model is running in the background, the web User Interface (UI) will show an animation to indicate that the process is still running.

\textbf{Fourth}, the generated title is displayed in the \tool's web UI, and is ready to be copied to GitHub's new pull request page.

\subsection{Deployment}
We provide a Dockerfile in our replication package\footnote{\url{https://github.com/soarsmu/Auto-PR-Title}} for easy deployment.
Users only need to build the image and run it as a container on every machine that supports Docker usage.
This practice will enable non-Python developers to use our apps, as no Python knowledge (i.e., environment set up, execution, etc.) is needed to deploy \tool in their machines.
\section{Usage Scenario}

\begin{figure}[t]
\centering
\includegraphics[width=0.45\textwidth]{img/US-1.pdf}
\caption{The scenario of using \tool}
\label{fig:flow}
\end{figure}

\begin{figure}[t]
\centering
\includegraphics[width=0.5\textwidth]{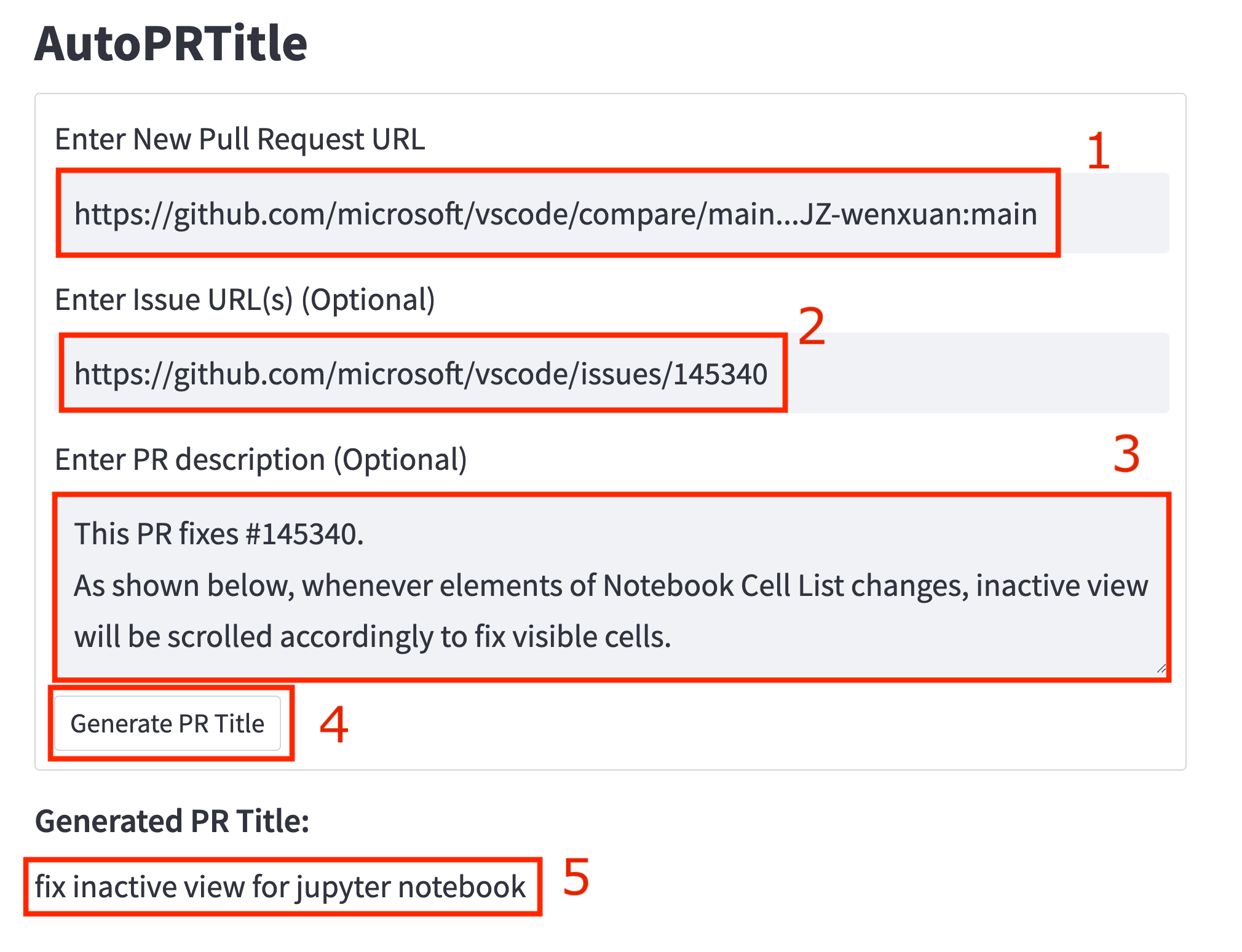}
\caption{The user interface of \tool}
\label{fig:ui}
\end{figure}

We illustrate how a user can use \tool in Figure~\ref{fig:flow}. 
We also show the user interface of \tool in Figure~\ref{fig:ui}.
Firstly, a user opens a new \pr page on GitHub: they can open a \pr page in a repository and click the \textit{New Pull Request} button.
Secondly, the user needs to copy and paste the new \pr URL to an input field (the red box 1 in Figure~\ref{fig:ui}).
They can then put the issue links, which are resolved by the new \pr, to another input field (the red box 2 in Figure~\ref{fig:ui}).
Furthermore, they can provide a description of this \pr on another input field (the red box 3 in Figure~\ref{fig:ui}).
Since the new pull request URL is the only mandatory input, the user can leave the other inputs empty.
After clicking \textit{Generate PR Title} (the red box 4 in Figure~\ref{fig:ui}), the generated \pr title will be output below it (the red box 5 in Figure~\ref{fig:ui}). The user can copy this generated title to the new pull request page and submit the \pr. 

In Figure~\ref{fig:ui}, we show an example of the generated \pr title produced by \tool on an existing \pr in \texttt{Microsoft/vscode} repository.\footnote{The \pr is available at \url{https://github.com/microsoft/vscode/pull/146125}}
This example contains all three input sources.
The original \pr title is \texttt{Fixes \#145340}.
This title is not informative and asks for an extra effort from \pr reviewer to check what the issue is actually about.
The \pr title generated by \tool is \texttt{fix inactive view for jupyter notebook}, which summarized the related commit messages and the linked issue title.
\tool can save \pr reviewers' time since they can understand the purpose of the \pr just by reading the generated \pr title.
\section{Evaluation}
\subsection{Dataset}
We built a dataset to facilitate automatic \pr title generation.
We considered a diverse set of repositories.
We first collected \pr from the Top-100 most-starred, Top-100 most-forked GitHub repositories regardless of programming languages.
We further included the Top-100 most-starred repositories that are written primarily in any of the following programming languages: JavaScript, Python, Java, C, and C++.
In total, we have a list of 700 repositories.
After filtering out the duplicate repositories, we are left with 578 distinct repositories.
We crawled the \pr of these repositories that were published before the year of 2022.
We then cleaned the dataset, and 83 distinct repositories were removed because all the corresponding PRs were filtered out.
The detailed steps can be found in our research paper~\cite{prtiger2022-arxiv}.
In the final dataset, we have 43,816 \pr from 495 distinct GitHub repositories.
We split them in the ratio of 8:1:1 for training, validation, and testing purposes.

\subsection{Evaluation Setup and Results}
\begin{table}[t]
\caption{Automatic evaluation results on the test dataset}
\centering
\begin{tabular}{@{}lcrrr@{}}
\toprule
\textbf{Approach} &
\textbf{ROUGE-1} & 
\textbf{ROUGE-2} &
\textbf{ROUGE-L} \\
\midrule
\textbf{BART}~\cite{lewis2019bart} & \textbf{47.22} & \textbf{25.27} & \textbf{43.12} \\
\textbf{PRSummarizer}~\cite{liu2019automatic} & 37.91 & 17.99 & 34.98 \\
\textbf{iTAPE}~\cite{chen2020stay} & 32.23 & 12.91 & 29.31 \\
\bottomrule
\end{tabular}
\label{tab:result}
\end{table}

We conducted two sets of evaluation: (1) {\em Model accuracy}: we calculated the ROUGE scores~\cite{lin2004rouge} of the generated \pr titles with the original high-quality \pr titles. (2) {\em Tool usability}: we invited evaluators to evaluate our tool usability.

\vspace{0.1cm} \noindent \textbf{(1) Model accuracy.} We report ROUGE-N (N=1,2) and ROUGE-L F1-scores. 
ROUGE-1 and ROUGE-2 measure the overlap of uni-grams (1-grams) and bi-grams (2-grams), respectively.
On the other hand, ROUGE-L measures the longest common sub-sequence between the reference summary (i.e., the original \pr title) and the generated summary (i.e., the model generated \pr title).
The formulas to calculate ROUGE-N (N=1,2) scores are as follows:
\begin{equation*}
    R_{rouge-n} = \frac{Count(overlapped\_N\_grams)}{Count(N\_grams \in reference\_summary)}
\end{equation*}
\begin{equation*}
    P_{rouge-n} = \frac{Count(overlapped\_N\_grams)}{Count(N\_grams \in generated\_summary)}
\end{equation*}
\begin{equation*}
    F1_{rouge-n} = 2\times \frac{R_{rouge-n}\times P_{rouge-n}}{R_{rouge-n}+ P_{rouge-n}}
\end{equation*}

\noindent Note that $overlapped\_N\_grams$ refers to the n-grams that appear in both the reference summary and the generated summary.
ROUGE F1-score ($F1_{rouge-n}$) seeks the balance between ROUGE Recall ($R_{rouge-n}$) and ROUGE Precision ($P_{rouge-n}$).
Thus, we use ROUGE F1-scores as the main metrics in our automatic evaluation.

\textbf{Result.} Table~\ref{tab:result} shows the model accuracy.
It indicates that BART outperforms both prior approaches.
BART outperforms PRSummarizer, which is a \pr description generation approach, by 24.6\%, 40.5\%, and 23.3\% in terms of ROUGE-1, ROUGE-2, and ROUGE-L F1-score, respectively.

\vspace{0.1cm}\noindent\textbf{(2) Tool usability.} We invited 7 people from our research group.
They all have at least 5 years of experience in programming and at least 4 years of experience using GitHub.
We give them access to our tool.
We also provided an example on how to use our tool.
However, we did not provide a list of existing pull requests for them. They are free to test our tool with any \pr in public repositories.
The rationale is that they can select repositories which they are more familiar with so that they can better judge the quality of the recommended \pr title.
We then asked them to provide two scores regarding the tool's usability.
Each score ranges from 1 to 5, indicating strongly disagree, slightly disagree, neutral, slightly agree, and strongly agree.
The two scores relate to two aspects, i.e., (1) \textit{Ease-of-use:} the tool is easy to use; (2) \textit{Usefulness:} The tool is useful and can help me to come up with a good \pr title.

\textbf{Result.} The averaged score in terms of \textit{ease-of-use} is 4.3.
While the average score about \textit{usefulness} is 4.4.
It shows that the evaluators perceive our tool to be useful and easy to use.
\section{Related Work}
Our work belongs to the broader research topic of automatically generating software artifacts.
In the past years, various tools have been proposed to generate different types of artifacts automatically.
For instance, AutoComment was proposed to automatically generate source code comments by leveraging Stack Overflow posts~\cite{wong2013autocomment}.
These automatic approaches, especially the ones related to summarization, are typically aimed to improve the quality of software artifacts and save practitioners' time~\cite{chen2020stay}.
Among the existing summarization tasks, the two most similar tasks to ours are Stack Overflow post title generation and issue title generation.
\textsc{Code2Que}~\cite{gao2021code2que} introduced a tool to generate a Stack Overflow post title based on the given code snippet.
\textsc{Code2Que} was intended to help developers write high-quality Stack Overflow question titles.
Besides generating titles, \textsc{Code2Que} also recommends related Stack Overflow questions based on the similarity between code snippets.
The other related tool is \textsc{iTiger}~\cite{itiger2022-arxiv}, which is aimed to automatically generate an issue title based on the given issue description.
\textsc{iTiger} has been implemented as a Userscript and provides the suggested title directly on GitHub's new issue page.
\tool differs from these tools as it requires several input sources: a \pr description, commit messages, and linked issue titles.
In comparison, the two relevant tools only require one type of input, i.e., code snippets or issue descriptions.
Thus, we implemented \tool as a web application to better capture the different sources of information in a \pr.
\section{Conclusion and Future Work}
In this work, we demonstrate an automatic \pr title generation tool, \tool.
Upon creating a new \pr, developers need to copy and paste the new \pr URL into our web application.
They can also add the associated issue links and a \pr description to the web application.
After clicking on the \textit{Generate PR Title}, our tool will return a suggested title.
Developers can then copy this title back to the new \pr page and submit it. \tool uses BART, a state-of-the-art summarization approach. In our evaluation, BART outperforms existing baselines. 
Moreover, evaluators agree that \tool is easy to use and useful in generating good \pr titles.
In the future, we would like to improve \tool by leveraging more types of information, such as code comments in the code changes.

\section*{Acknowledgment}
This research / project is supported by the National Research Foundation, Singapore, under its Industry Alignment Fund – Pre-positioning (IAF-PP) Funding Initiative. Any opinions, findings and conclusions or recommendations expressed in this material are those of the author(s) and do not reflect the views of National Research Foundation, Singapore.

\balance

\bibliographystyle{IEEEtran}
\bibliography{main}

\begin{thebibliography}{10}
\providecommand{\url}[1]{#1}
\csname url@samestyle\endcsname
\providecommand{\newblock}{\relax}
\providecommand{\bibinfo}[2]{#2}
\providecommand{\BIBentrySTDinterwordspacing}{\spaceskip=0pt\relax}
\providecommand{\BIBentryALTinterwordstretchfactor}{4}
\providecommand{\BIBentryALTinterwordspacing}{\spaceskip=\fontdimen2\font plus
\BIBentryALTinterwordstretchfactor\fontdimen3\font minus
  \fontdimen4\font\relax}
\providecommand{\BIBforeignlanguage}[2]{{%
\expandafter\ifx\csname l@#1\endcsname\relax
\typeout{** WARNING: IEEEtran.bst: No hyphenation pattern has been}%
\typeout{** loaded for the language `#1'. Using the pattern for}%
\typeout{** the default language instead.}%
\else
\language=\csname l@#1\endcsname
\fi
#2}}
\providecommand{\BIBdecl}{\relax}
\BIBdecl

\bibitem{yu2015wait}
Y.~Yu, H.~Wang, V.~Filkov, P.~Devanbu, and B.~Vasilescu, ``Wait for it:
  Determinants of pull request evaluation latency on github,'' in \emph{2015
  IEEE/ACM 12th working conference on mining software repositories}.\hskip 1em
  plus 0.5em minus 0.4em\relax IEEE, 2015, pp. 367--371.

\bibitem{liu2019automatic}
Z.~Liu, X.~Xia, C.~Treude, D.~Lo, and S.~Li, ``Automatic generation of pull
  request descriptions,'' in \emph{2019 34th IEEE/ACM International Conference
  on Automated Software Engineering (ASE)}.\hskip 1em plus 0.5em minus
  0.4em\relax IEEE, 2019, pp. 176--188.

\bibitem{fang2022prhan}
S.~Fang, T.~Zhang, Y.-S. Tan, Z.~Xu, Z.-X. Yuan, and L.-Z. Meng, ``{PRHAN:}
  automated pull request description generation based on hybrid attention
  network,'' \emph{Journal of Systems and Software}, vol. 185, p. 111160, 2022.

\bibitem{prtiger2022-arxiv}
T.~Zhang, I.~C. Irsan, F.~Thung, D.~Han, D.~Lo, and L.~Jiang, ``Automatic pull
  request title generation,'' 2022 IEEE 38th International Conference on
  Software Maintenance and Evolution (ICSME), Research Track.

\bibitem{lewis2019bart}
M.~Lewis, Y.~Liu, N.~Goyal, M.~Ghazvininejad, A.~Mohamed, O.~Levy, V.~Stoyanov,
  and L.~Zettlemoyer, ``{BART:} denoising sequence-to-sequence pre-training for
  natural language generation, translation, and comprehension,'' in
  \emph{Proceedings of the 58th Annual Meeting of the Association for
  Computational Linguistics, {ACL} 2020, Online, July 5-10, 2020}.\hskip 1em
  plus 0.5em minus 0.4em\relax Association for Computational Linguistics, 2020,
  pp. 7871--7880.

\bibitem{chen2020stay}
S.~Chen, X.~Xie, B.~Yin, Y.~Ji, L.~Chen, and B.~Xu, ``Stay professional and
  efficient: Automatically generate titles for your bug reports,'' in
  \emph{2020 35th IEEE/ACM International Conference on Automated Software
  Engineering (ASE)}.\hskip 1em plus 0.5em minus 0.4em\relax IEEE, 2020, pp.
  385--397.

\bibitem{zt2022benchmark}
T.~Zhang, D.~P. Chandrasekaran, F.~Thung, and D.~Lo, ``Benchmarking library
  recognition in tweets,'' in \emph{2022 IEEE/ACM 30th International Conference
  on Program Comprehension (ICPC)}, 2022, pp. 343--353.

\bibitem{zhang2020sentiment}
T.~Zhang, B.~Xu, F.~Thung, S.~A. Haryono, D.~Lo, and L.~Jiang, ``Sentiment
  analysis for software engineering: How far can pre-trained transformer models
  go?'' in \emph{2020 IEEE International Conference on Software Maintenance and
  Evolution (ICSME)}.\hskip 1em plus 0.5em minus 0.4em\relax IEEE, 2020, pp.
  70--80.

\bibitem{vaswani2017attention}
A.~Vaswani, N.~Shazeer, N.~Parmar, J.~Uszkoreit, L.~Jones, A.~N. Gomez,
  {\L}.~Kaiser, and I.~Polosukhin, ``Attention is all you need,''
  \emph{Advances in neural information processing systems}, vol.~30, 2017.

\bibitem{liu2019roberta}
Y.~Liu, M.~Ott, N.~Goyal, J.~Du, M.~Joshi, D.~Chen, O.~Levy, M.~Lewis,
  L.~Zettlemoyer, and V.~Stoyanov, ``Roberta: A robustly optimized bert
  pretraining approach,'' \emph{arXiv preprint arXiv:1907.11692}, 2019.

\bibitem{dinan2020second}
E.~Dinan, V.~Logacheva, V.~Malykh, A.~Miller, K.~Shuster, J.~Urbanek, D.~Kiela,
  A.~Szlam, I.~Serban, R.~Lowe \emph{et~al.}, ``The second conversational
  intelligence challenge (convai2),'' in \emph{The NeurIPS'18
  Competition}.\hskip 1em plus 0.5em minus 0.4em\relax Springer, 2020, pp.
  187--208.

\bibitem{see2017get}
A.~See, P.~J. Liu, and C.~D. Manning, ``Get to the point: Summarization with
  pointer-generator networks,'' in \emph{Proceedings of the 55th Annual Meeting
  of the Association for Computational Linguistics, {ACL} 2017, Vancouver,
  Canada, July 30 - August 4, Volume 1: Long Papers}, R.~Barzilay and M.~Kan,
  Eds.\hskip 1em plus 0.5em minus 0.4em\relax Association for Computational
  Linguistics, 2017, pp. 1073--1083.

\bibitem{gu2016incorporating}
J.~Gu, Z.~Lu, H.~Li, and V.~O.~K. Li, ``Incorporating copying mechanism in
  sequence-to-sequence learning,'' in \emph{Proceedings of the 54th Annual
  Meeting of the Association for Computational Linguistics, {ACL} 2016, August
  7-12, 2016, Berlin, Germany, Volume 1: Long Papers}.\hskip 1em plus 0.5em
  minus 0.4em\relax The Association for Computer Linguistics, 2016.

\bibitem{Streamlit}
``Streamlit,'' \url{http://streamlit.io}, (Accessed on 06/23/2022).

\bibitem{lin2004rouge}
C.-Y. Lin, ``Rouge: A package for automatic evaluation of summaries,'' in
  \emph{Text summarization branches out}, 2004, pp. 74--81.

\bibitem{wong2013autocomment}
E.~Wong, J.~Yang, and L.~Tan, ``Autocomment: Mining question and answer sites
  for automatic comment generation,'' in \emph{2013 28th IEEE/ACM International
  Conference on Automated Software Engineering (ASE)}.\hskip 1em plus 0.5em
  minus 0.4em\relax IEEE, 2013, pp. 562--567.

\bibitem{gao2021code2que}
Z.~Gao, X.~Xia, D.~Lo, J.~Grundy, and Y.-F. Li, ``{Code2Que:} a tool for
  improving question titles from mined code snippets in stack overflow,'' in
  \emph{Proceedings of the 29th ACM Joint Meeting on European Software
  Engineering Conference and Symposium on the Foundations of Software
  Engineering}, 2021, pp. 1525--1529.

\bibitem{itiger2022-arxiv}
\BIBentryALTinterwordspacing
T.~Zhang, I.~C. Irsan, F.~Thung, D.~Han, D.~Lo, and L.~Jiang, ``{iTiger:} an
  automatic issue title generation tool,'' 2022. [Online]. Available:
  \url{https://arxiv.org/abs/2206.10811}
\BIBentrySTDinterwordspacing

\end{thebibliography}

\end{document}